\documentclass[12pt]{article}
\usepackage{geometry}
\usepackage{booktabs}
\usepackage{url,bm,amsmath,graphicx,amsfonts}
\usepackage{color}
\usepackage{natbib}
\usepackage[colorlinks, linkcolor=blue, citecolor=blue]{hyperref}

\usepackage{setspace}
\usepackage{caption}
\newcommand{\dd}{\mathrm{d}}

\renewcommand{\hat}{\widehat}
\renewcommand{\bar}{\overline}
\renewcommand{\tilde}{\widetilde}

\newtheorem{Proposition}{Proposition}

\title{Bayesian Clustered Coefficients Regression with Auxiliary Covariates
Assistant Random Effects}

\author{Guanyu Hu~~Yishu Xue~~Zhihua Ma\thanks{Correspondence:
mazh1993@outlook.com}}

\date{}

\begin{document}
\maketitle
\begin{abstract}

In regional economics research, a problem of interest is to detect similarities
between regions, and estimate their shared coefficients in economics models. In
this article, we propose a mixture of finite mixtures (MFM) clustered
regression
model with auxiliary covariates that account for similarities in demographic or
economic characteristics over a spatial domain. Our Bayesian construction
provides both inference for number of clusters and clustering configurations,
and estimation for parameters for each cluster. Empirical performance of the
proposed model is illustrated through simulation experiments, and further
applied to a study of influential factors for monthly housing cost in Georgia.
\bigskip 

\end{abstract}

\section{Introduction}\label{sec:intro}
Analysis of spatial data referenced over different locations has received
widespread attention in many fields such as environmental science
\citep{hu2018stat,yang2019bayesian}, social science
\citep{bradley2018computationally}, and biostatistics \citep{xu2019latent}.
There are two major approaches to analyzing spatial data. The first approach,
where spatial variations in the outcome are accounted for by an additive
spatial
random effect term at each location, has been studied both for the linear model
\citep{cressie1992statistics} and generalized linear model
\citep{diggle1998model} settings. The second approach, a regression model with
spatially varying coefficients, is developed to capture spatial variations
within the covariate effects themselves. For the spatially varying coefficients
model, there are two major approaches for estimation of the regression
coefficients. One is geographically weighted regression
\citep[GWR;][]{brunsdon1996geographically}, the basic idea of which is to
assign
different weights to the observations based on a certain measure of distance
between them and the target location. This work also has various extensions in
generalized linear regression \citep{nakaya2005geographically} and analysis of
survival data \citep{hu2018modified,xue2019geographically},
as well as under the Bayesian paradigm \citep{ma2019bayesian}.
It has been further extended to multiscale models that allow kernels of
different variables to be on
different scales, which greatly enhances its model flexibility
\citep{fotheringham2017multiscale}. Another major approach is to give a
Gaussian
process prior to the spatially varying coefficients \citep{gelfand2003spatial},
which provides a natural and flexible way to view the coefficient surface as a
realization from a spatial process. This work is universally applied into
different models such as Poisson regression model \citep{reich2010bayesian} and
survival models \citep{hu2020comparison}.

Most recently, heterogeneous covariate effects in many different fields, such
as
real estate applications, spatial econometrics, and environmental science are
receiving increasing attention. For example, a country's big cities and small
cities could be put into separate clusters and analyzed, as each subgroup share
more similarities in development patterns, and thus similar covariate effects
in
regression models can be expected.
Such clustering information is of great interest to regional economics
researchers. Existing frequentist approaches include those based on scan
statistics \citep{kulldorff1995spatial}, two-step spatial hypothesis testing
\citep{lee2017cluster,leespatial2019}, and a penalized method based on minimum
spanning tree \citep{li2019spatial}.
Under the Bayesian paradigm, an integrated framework
\citep{ma2019bayesianspatial} to detect clusters in the covariate effects as
well as producing the parameter estimates for spatially dependent data are
proposed, in which clustering is done via the Dirichlet process mixture model
\citep[DPM;][]{neal2000markov,ishwaran2002exact}. The DPM, however, has turned
out to
produce extremely small clusters and make the estimation for number of clusters
inconsistent \citep{miller2013simple}. The mixture of finite mixture (MFM)
model
proposed by \cite{miller2018mixture} provides a remedy to over-clustering
problem for Bayesian nonparametric methods. Both DPM and MFM allow for
uncertainty in the number of clusters instead of relying on a given number,
which needs to be tuned based on certain criteria.

While spatial random effects have been used to account for the influence of
geographical proximity on the similarity of outcomes for two neighboring
observations, in existing approaches, the correlation between spatial random
effect terms only depends on distance, and all other factors are ignored.
For improvement in describing spatial correlation, some works have been done to
bring in auxiliary information to help the estimation of spatial regression
model \citep{white2009stochastic,lee2014bayesian,gao2019bayesian}. These works
focus on the estimation of the edge based on some covariates or spatial
structure. The regression relationship between the $p$-dimensional covariance
matrix and auxiliary information has been explored by \cite{zou2017covariance}
and \cite{liu2020semiparametric}, which can help reveal the true correlation
structure of spatial data. Given the importance of spatial random effects in a
spatial regression model, their covariance structure need to be appropriately
specified.

In this work, we propose a Bayesian clustered linear regression model with MFM,
which when compared to the DPM, consistently estimates the number of clusters.
The estimation of parameters remains precise. To our best knowledge, we firstly
introduce the MFM in clustered coefficients regression model. In
addition,
a weighted average correlation structure of auxiliary covariates information is
incorporated in linear mixed effects regression model based on Dirichlet
prior.


The remainder of the paper is organized as follows. In
Section~\ref{sec:method},
we present the spatial clustered linear regression model with MFM. In
Section~\ref{sec:Infer} , a MCMC sampling algorithm based on \textbf{nimble}
\citep{de2017programming} and post MCMC inference are discussed. Extensive
simulation studies are performed in Section~\ref{sec:simu}. For illustration,
our proposed methodology is applied to Georgia housing cost data in
Section~\ref{sec:real_data}. We conclude this paper with a brief discussion in
Section~\ref{sec:discussion}.

\section{Method}\label{sec:method}
\subsection{Spatial Linear Regression}

The basic geostatistical model \citep[][]{gelfand2016spatial}
for observations made over a spatial domain can be written as:
\begin{equation}
\label{eq:spatial_regression}
\bm{Y} = \bm{X}\bm{\beta} + \bm{w} + \bm{\epsilon},
\end{equation}
where $\bm{Y} = (Y(s_1),\ldots, Y(s_n))$ denotes the $n$-dimensional vector of
spatial responses observed at locations $s_1,\ldots,s_n$, $\bm{X} =
\begin{pmatrix} X(s_1)^\top \\ \ldots \\ X(s_n)^\top \end{pmatrix}$ denotes
the~$n\times p$ matrix of covariates, $\bm{\beta}$ is the vector of
coefficients, $\bm{w} = (w(s_1),\ldots, w(s_n))^\top$ is a vector of spatial
random effects, and $\bm{\epsilon}\sim \mbox{MVN}(\bm{0}, \tau_y^{-1}\bm{I})$
is
the ``nugget effect'' with $\tau_y$ being the precision of the response
\citep[Chapter~6,][]{carlin2014hierarchical}. The above spatial regression
model
can be formulated alternatively as
\begin{equation*}
\label{eq:spatial-regression-2}
\begin{split}
Y(s_i) \mid \bm{\beta}, w(s_i), \tau_y &\sim \mbox{N}(\bm{X}(s_i)\bm{\beta}
+ w(s_i),
\tau_y^{-1}\bm{I}), ~~i=1,\ldots, n,\\
\bm{w} &\sim \mbox{MVN}(\bm{0},\bm{\Sigma}_{\bm{w}}),
\end{split}
\end{equation*}
where~$\bm{w}$ denotes the spatial structure with a covariance
matrix~$\bm{\Sigma}_{\bm{w}}$, and~N and MVN denote the univariate and
multivariate normal distributions, respectively. The covariance
matrix~$\bm{\Sigma_w}$ is often defined to be~$\sigma_w^2\bm{H}$, with~$\bm{H}$
constructed using the great circle distance (GCD) matrix among different
locations via the following three popular weighting schemes:
\begin{equation}\label{eq:weightingscheme}
\begin{split}
\text{the unity scheme}: & ~~\bm{H} = \text{diag}(1) \\
\text{the exponential scheme}: & ~~\bm{H} = \exp(-\mbox{GCD} / \phi) \\
\text{the Gaussian scheme}: & ~~\bm{H} = \exp(-(\mbox{GCD}/\phi)^2),
\end{split}
\end{equation}
where $\phi$ is a bandwidth parameter that controls the spatial correlation.

Instead of using spatial random effects, i.e., location-wise intercepts
to account for spatial variations in $\bm{Y}$,
the
spatially varying coefficients model \citep[][]{gelfand2003spatial} attributes
such effects to variations in the parameters over the spatial domain, i.e., the
parameter~$\bm{\beta}$ itself varies. Such a model is formulated as:
\begin{eqnarray}
Y(s_i)=\bm{X}^\top (s_i)\tilde{\bm{\beta}}(s_i)+\epsilon(s_i),
\label{svc model}
\end{eqnarray}
where $\bm{X}(s_i)$ is vector of covariates at location of the $i$th subject
$s_i$, and~$\tilde{\bm{\beta}}(s_i)$ is assumed to be generated from
a~$p$-variate spatial process model. With observations $(Y(s_i),\bm{X}(s_i))$
for $i=1,\ldots, n$, the model can be written as 
\begin{equation*}
\bm{Y} = \bm{X}^\top \tilde{\bm{\beta}} + \bm{\epsilon},
\end{equation*} 
where $\bm{Y}=(Y(s_1), \ldots, Y(s_n))^\top$, $\bm{X}^\top$ is an $n\times
(np)$
block diagonal matrix whose $i$-th diagonal entry is $\bm{X}^\top(s_i)$,
$\tilde{\bm{\beta}} = (\tilde{\bm{\beta}}(s_1)^\top, \ldots,
\tilde{\bm{\beta}}(s_n)^\top)^\top$, and $\bm{\epsilon}\sim
\mbox{MVN}(0,\tau_y^{-1}\bm{I})$ with~$\bm{I}$ being the identity matrix. To
characterize the~$p$-variate spatial process that
generates~$\tilde{\bm{\beta}}(s_i)$, we rewrite the model as
\begin{equation}
\begin{split}
\bm{Y} \mid \tilde{\bm{\beta}},\tau_y^{-1} & \sim
\mbox{MVN}(\bm{X}^\top\tilde{\bm{\beta}},\tau_y^{-1}\bm{I}), \\
{\tilde{\bm{\beta}}}\mid \bm{\mu}_{\bm{\beta}},\bm{T} & \sim
\mbox{MVN}(1_{n\times 1}\otimes
\bm{\mu}_{\bm{\beta}},\bm{H}(\phi)\otimes \bm{T}),
\end{split}
\label{hierarchical model2}
\end{equation}
where $\bm{\mu}_{\bm{\beta}}$ is a $p \times 1$ vector,  $\bm{H}(\phi)$ is an
$n\times n$-dimensional matrix measuring spatial correlations between the~$n$
observed locations, $\bm{T}$ is a $p \times p$ covariance matrix associated
with
an observation vector at any spatial location, and~$\otimes$ denotes the
Kronecker product.


In the above formulation, each location has its own~$\bm{\beta}$ vector.
However, such a model could be too flexible, as there are certain regions
that have very similar~$\bm{\beta}$ values.
From a modeling perspective, clustering such
regions and having them all share one parameter vector encourages a
parsimonious model without compromising the model's explanatory power.
Another potential drawback of formulation in (\ref{hierarchical
model2}) is that all variations are accounted for by~$\tilde{\bm{\beta}}$,
and spatial random effects are ignored. However, the random effects term is
rather important, and influenced not only by distance but also other factors
such as demographics, transportation, etc. Therefore, a model with clustered
coefficients, and also random effects terms that help account for the
intricate connections between regions is desired.

\subsection{Mixture of Finite Mixture Model} \label{sec:MFM}

Based on the heterogeneity pattern, we focus on the clustering of
spatially-varying coefficients. A latent clustering structure can be introduced
to accommodate the spatial heterogeneity on parameters of sub-areas. Let us
denote the cluster belongings of the~$i$th observation as $z_i$
for~$i=1,\ldots,
n$. The Dirichlet process \citep[DP;][]{ferguson1973bayesian} offers a
nonparametric approach for capturing heterogeneity effects in the data. The DP
prior for the cluster belonging of the $i$th observation can be written as
\begin{equation}
\begin{split}
z_i \mid k, \pi & \sim \sum_{h=1}^k  \pi_h \delta_{h}, \quad  i=1, \ldots, n,\\
(\pi_1, \ldots, \pi_k) \mid k &\sim \text{Dirichlet}(\alpha/k, \ldots,
\alpha/k),
\end{split}
\label{eq:dpmm}
\end{equation}
where $k \rightarrow \infty$, $\pi_h$ denotes the random probability weight,
$\delta_h$ is the Dirac~$\delta$ with point mass at~$h$, and $\alpha$ is the
concentration parameter.
The first equation in \eqref{eq:dpmm} can be expressed equivalently as a
multinomial distribution:
\begin{equation*}
	P(z_i = h \mid k,\pi) = \begin{cases}
	\pi_1, & h=1 \\
	\vdots \\
	\pi_k, & h = k
\end{cases},
\end{equation*}
and we will use the~$\sum_{h=1}^k  \pi_h \delta_{h}$ notation for simplicity
for the rest of this paper.
The joint distribution for $z_1,\cdots,z_n$ can also
be
written as a conditional distribution, known as the Chinese restaurant process
\citep[CRP;][]{pitman1995exchangeable, neal2000markov}. The distribution
of~$z_i$ is marginally represented by the stick-breaking construction of
\cite{sethuraman1994constructive} as
\begin{equation}
	\begin{split}
		z_i &\sim \sum_{h=1}^k \pi_h\delta_h,\\
		\pi_h&=\zeta_h\prod_{\ell<h}(1-\zeta_l),\\
		\zeta_h&\sim \text{Beta}(1,\alpha).
	\end{split}
	\label{eq:dp_sb}
\end{equation}
\cite{miller2013simple} showed that the posterior distribution on the number of
clusters does not converge to the true number of components, and extraneous
clusters are often produced by the CRP. Later,
\cite{miller2018mixture}
proposed a modification of the CRP called a mixture of finite mixtures (MFM)
model to circumvent this issue, which can be formulated as
\begin{eqnarray}\label{eq:MFM}
\begin{split}
k & \sim p(\cdot), \\
(\pi_1, \ldots, \pi_k) \mid k &\sim \text{Dirichlet}(\gamma, \ldots, \gamma),
\\
 z_i \mid k, \pi & \sim \sum_{h=1}^k  \pi_h \delta_h, \quad  i=1, \ldots, n, 
\end{split}
\end{eqnarray}
where~$p(\cdot)$ is a proper probability mass function (p.m.f) on~$\{1, 2,
\ldots\}$. A default choice of $p(\cdot)$ is a $\mbox{Poisson}(1)$ distribution
truncated to be positive \citep{miller2018mixture}, which is assumed through
the
rest of the paper. Analogous to the stick-breaking representation
in~\eqref{eq:dp_sb}, the MFM also has a similar construction. If we choose $k-1
\sim \mbox{Poisson}(\lambda)$ and $\gamma=1$ in \eqref{eq:MFM}, the mixture
weights~$\pi_1,\cdots,\pi_k$ can be constructed as:
\begin{enumerate}
	\item Generate $\eta_1,\eta_2,\cdots \overset{\text{iid}}{\sim}
\text{Exp}(\lambda)$,
	\item $k=\min\{j:\sum_{i=1}^j \eta_i\geq 1\}$,
	\item $\pi_i=\eta_i$, for $i=1,\cdots,k-1$,
	\item $\pi_k=1-\sum_{i}^{k-1}\pi_i$.
\end{enumerate}
Gibbs samplers are easily constructed in stick-breaking framework
\citep{ishwaran2001gibbs}. For ease of exposition, we refer to the
formulation in~\eqref{eq:MFM} as $\text{MFM}(\gamma,\lambda)$.

\subsection{Auxiliary Covariates Assistant Covariance Matrix}

In the regression model \eqref{eq:spatial_regression} where spatial random
effects
are present, their covariance structure often depend on the geographical
distance between pairs of locations, which in general indicates that closer
locations have stronger correlation, and is in accordance with Tobler's first
law of geography that ``everything is related to everything else, but near
things are more related than distant things''. In most economics problems,
however, spatial proximity might not be the sole indicator for similarity, as
there can be geographically distant locations that share similar demographical
characteristics. For example, while the GCD between New York City
and Albany is only~135 miles, which is far less than the~2569 miles between
New York City and San Francisco (calculated using \textbf{ggmap} and
\textbf{geosphere} packages in \textsf{R}), the population density of Albany is
only 4525.3 per square mile, which is far smaller than those of New York City
and San Francisco, which are, respectively, 27,709.4 and~19,104.4 per square
mile \citep[Data source:][]{worldpopulation2020}.
To incorporate
such similarities into the covariance structure for random effects, motivated
by
covariance regression \citep{zou2017covariance} and Bayesian model averaging
\citep{raftery1997bayesian}, we propose the following auxiliary covariates
assistant covariance (ACAC) matrix for random effects in a mixture regression
model:
\begin{equation}
\begin{split}
	\bm{w} &\sim \text{MVN}(\bm{0},\bm{\Sigma}_{\bm{w}}),\\
	\bm{\Sigma}_{\bm{w}}&=\sigma^2\left\{\alpha_0\bm{I}_n+\alpha_1\bm{W}
	(\bm{Z}_1)+\cdots+\alpha_J\bm
{W}(\bm{Z}_J)\right\},
\end{split}
\label{eq:covariance_regression}
\end{equation} 
where $\sigma^2$ is a constant accounting for the overall magnitude of
variance,
$\bm{w}=(w(s_1),\cdots,w(s_n))^\top$, and $\bm{W}(\bm{Z}_j),j=1,\cdots,J$ is
the
similarity matrix of the $j$th auxiliary covariate. Entries of the similarity
matrix have values between~0 and~1, and are usually decreasing with respect to
the absolute difference between values of the auxiliary covariates. The three
aforementioned weighting schemes in~\eqref{eq:weightingscheme} can be used to
define~$\bm{W}(\bm{Z}_j)$. For example, an exponential decay similarity
matrix~$\bm{W}(\bm{Z}_j)$ can be constructed so that its~$(\ell, \ell^{'})$-th
element is
\begin{equation} \label{eq:W}
	\exp(-\kappa_j|Z_j(\bm{s}_{\ell})-Z_j(\bm{s}_{\ell^{'}})|),
\end{equation}
where $\kappa_j>0$ is the range parameter for the
exponential kernel, and $|\cdot|$ denotes the Euclidean distance. In
order to solve the identifiability issue, we set a constraint for
$\alpha_0,\alpha_1,\cdots,\alpha_J$:
\begin{equation}
	\begin{split}
		\sum_{j=0}^J \alpha_j&=1,\\
		0\leq \alpha_j \leq 1,\, j&=0,1,\cdots,J.
	\end{split}
	\label{eq:alpha_constraints}
\end{equation} 

\begin{Proposition}
	If we have $J$ $n\times n$ positive definite matrices
	$\bm{\Sigma}_1,\ldots,$ $\bm{\Sigma}_J$,
	and a sequence of positive numbers $\alpha_0,\ldots,\alpha_J$
	which satisfy $\sum_{j=0}^J \alpha_j=1$ and $0\leq \alpha_j \leq 1$ for 
	$j=0,1,\cdots,J$, then the matrix $\bm{\Sigma}=\alpha_0\bm{I}_n+
	\sum_{i=1}^J \alpha_i\bm{\Sigma}_i$ is positive
	definite.
\end{Proposition}
Proof for this proposition is directed to Supplemental Section~S.1.

Based on the constraint in \eqref{eq:alpha_constraints}, a Dirichlet prior is
assigned to $\alpha_0,\cdots,$ $\alpha_J$. The prior distribution of
$\alpha_0,\alpha_1,\cdots,\alpha_J$ is given as
\begin{equation}
\alpha_0,\alpha_1,\cdots,\alpha_J \sim \text{Dirichlet}(\nu),
\label{eq:dir_prior}
\end{equation}
where $\text{Dirichlet}(\nu)$ is the Dirichlet distribution with parameter
$\nu$.

\subsection{Spatial MFM Clustered Regression with ACAC}
Combining the MFM and ACAC matrix, we have
our final spatial MFM clustered regression model with ACAC
hierarchically
as follows, for $i = 1, \cdots, n$:
\begin{equation}
	\begin{split}
	Y(s_i) \mid w(s_i),\bm{\beta}_{z_i},\tau_y
	&\stackrel{\text{ind}}{\sim}\mbox{N}(\bm{X}(s_i)
	\bm{\beta}_{z_i}+w(s_i), \tau_y^{-1}),\\
\bm{w} \mid \bm{\Sigma}_{\bm{w}} & \sim
\text{MVN}(\bm{0},\bm{\Sigma}_{\bm{w}}),\\
\bm{\Sigma}_{\bm{w}} &=\sigma^2\{\alpha_0 \bm{I}_n+\alpha_1\bm{W}(\bm{Z}_1)
+\cdots+\alpha_J\bm{W
}(\bm{Z}_J)\},\\
	\alpha_0,\alpha_1,\cdots,\alpha_J & \sim \text{Dirichlet}(\nu),\\
\beta_{z_i\ell} &\stackrel{\text{ind}} \sim \mbox{N}(\mu_{z_i\ell},
\tau_\beta^{-1}
), \ell = 1,\cdots, p,\\
\mu_{z_{i}\ell} & \sim \mbox{N}(0,1), \\
z_i & \sim \text{MFM}(\gamma,\lambda),\\
	\tau_y & \sim \text{Gamma}(a_1,b_1),\\
	\tau_\beta
&\sim\mbox{Gamma}(c_1, c_2),\\
	\sigma^2 &\sim \text{InverseGamma}(a_2,b_2),
	\end{split}
	\label{eq: hierarchical}
\end{equation}	
where $\bm{w} = (w(s_1), \cdots, w(s_n))^\top$,
$\bm{\beta}_{z_i} = (\beta_{z_i1}, \cdots, \beta_{z_ip})^\top$ with
$p$ being the dimension of the covariates
$\bm{X}(s_i)$, $\bm{W}(\cdot)$ is the similarity matrix of the corresponding
auxiliary covariate defined in (\ref{eq:W}), and MFM the clustering method
introduced in Section \ref{sec:MFM}. The choice of hyperparameters will be
discussed in Section~\ref{sec:Infer}.

\section{Bayesian Inference}\label{sec:Infer}

\subsection{Bayesian Computation}

Let $\bm{\theta}= \{(\tau_y, \mu_{z_i\ell}, \tau_{z_i\ell}, \sigma^2, \lambda,
\kappa_j, \bm{\alpha}): i = 1,\cdots,n; \ell = 1,\cdots,p; j=1,\cdots,J\}$
denote the set of unknown parameters in the proposed model, and we assume that
they are independent \textit{a priori}. Therefore, we assign commonly used
priors for these parameters: $\tau_y \sim
\text{Gamma}(1,1)$, $\mu_{z_i\ell} \sim \mbox{N}(0, 1)$, $\tau_{z_i\ell}
\sim
\text{Gamma}(1,1)$,
$\sigma^2 \sim \mbox{InverseGamma(1, 1)}$,
$\lambda \sim \text{log-normal}(0, 1)$, $1/\kappa_j \sim
\text{Gamma}(1, 1)$, and $\bm{\alpha} \sim \text{Dirichlet}(1,$ $1, \cdots)$.
With the prior distributions specified above, the posterior distribution of
these unknown parameters based on the data $D = \{Y(s_i), \bm{X}(s_i),$
$\bm{Z}_j\}$ is given by
\begin{align*}\pi(\bm{\theta}\mid D) &\propto
L(\bm{\theta}\mid D)\pi(\bm{\theta})\\
& \propto \int_{\bm{w}} \bigg[\prod_{i=1}^{n} f(Y(s_i)\mid w(s_i),
\bm{\beta}_{z_i}, \tau_y)f(\bm{\beta}_{z_i})f(z_i) f(\bm{w}\mid
\bm{\Sigma}_{\bm{w}})
\dd \bm{w}\bigg]\pi(\bm{\theta}).
\end{align*}
The analytical form of the posterior distribution of $\bm{\theta}$ is
unavailable. Therefore, we employ the Markov chain Monte Carlo (MCMC) sampling
algorithm to sample from the posterior distribution, and then obtain the
posterior estimates for the unknown parameters. Computation is facilitated by
the \textbf{nimble} package in \textsf{R}, which uses syntax similar to
\textsf{WinBUGS} and \textsf{JAGS}, but generates \textsf{C++} code for faster
computation. With the \textbf{nimble} package, sampling algorithms for
the parameters are default samplers. For parameters $\bm{\alpha}$, a
random-walk Dirichlet sampler is used. For $\bm{\beta}_{z_i}$, a random-walk
block sampler is used. The same sampler is used for parameter $\bm{w}$. The
conjugate sampler is used for $\tau_y$ and a categorical sampler is used for
$z_i$, while for the other parameters, a random-walk sampler is applied.

\subsection{Posterior Inference and Diagnostic}

In the proposed spatial regression model, since the covariance matrix of the
spatial
random
effects can be constructed in different ways, including the unity, exponential,
and Gaussian weighting schemes in \eqref{eq:weightingscheme}, a model selection
criterion needs to be used for deciding which form of the covariance matrix is
the most suitable for the data. A commonly used Bayesian model selection
criterion, logarithm of the pseudo-marginal likelihood
\citep[LPML;][]{ibrahim2013bayesian} can be used for this purpose. The LPML can
be obtained through the conditional predictive ordinate (CPO) values, which are
the Bayesian estimates for the probability of observing~$Y_i$ in the future
after other observations are made.
Let $Y^*_{(-i)} = \{Y_j: j = 1, \cdots, i-1,
i+1, \cdots, n\}$ denote the observations with the $i$th subject response
removed. The CPO for the $i$th subject is defined as:
\begin{equation}
\label{eq:CPO}
\text{CPO}_i = \int f(y(s_i)\mid\bm{\beta}_{z_i},w(s_i),\tau_y,
\bm{\alpha})\pi(\bm{w}, \bm{\alpha},
\bm{\beta}_{z_i}, \tau_y \mid Y_{(-i)}) \dd (\bm{w},\bm{\alpha},
\bm{\beta}_{z_i},
\tau_y),
\end{equation}
where 
\begin{align*}
\pi(\bm{w},\bm{\alpha}, \bm{\beta}_{z_i}, \tau_y &\mid Y^*_{(-i)}) \\
&=
\frac{\prod_{j \ne i}
f(y(s_j)\mid \bm{\beta}_{z_j},w(s_j),\tau_y,\bm{\alpha})\pi(\bm{w},\bm{\alpha},
\bm{\beta}_{z_j},
	\tau_y\mid Y^*_{(-i)})}{c(Y^*_{(-i)})},
\end{align*}
with $c(Y^*_{(-i)})$ being the normalizing constant.
As discussed in \cite{chen2012monte}, $\mbox{CPO}_i$ is also called the
cross-validated predictive density, and Equation~\eqref{eq:CPO} is essentially
integrating the predictive distribution of~$y(s_i)$ given the rest of the
observations. It is a useful quantity for model checking, as it describes how
much the observation at location~$s_i$ supports the model. An equivalent
expression for~$\mbox{CPO}_i$ is:
\begin{equation}\label{eq:cpo_2}
\
\mbox{CPO}_i = \frac{1}{\int \frac{1}{f(y(s_i)\mid
\bm{\theta})}p(\bm{\theta\mid \bm{y}})\dd \bm{\theta}},
\end{equation}
where~$\bm{\theta}$ denote the parameters of the model.
Let $\{\bm{\theta}_t,~t=1,\ldots,T\}$ denote
a Gibbs sample of~$\bm{\theta}$ from~$p(\bm{\theta\mid \bm{y}})$, using
Equation~\eqref{eq:cpo_2},
a Monte Carlo estimate of the CPO can be obtained as:
\begin{equation}
\label{eq:CPOest}
\widehat{\text{CPO}}_i^{-1} = \frac{1}{T} \sum_{t=1}^{T}
\frac{1}{f(y(s_i)\mid w_t(s_i), \bm{\alpha}_t, \bm{\beta}_{z_i,t},
\tau_{y,t})},
\end{equation}
where $T$ is the total number of Monte Carlo iterations. Based on
$\hat{\mbox{CPO}_i}$, the LPML can be estimated as:
\begin{equation}
\label{eq:LPML}
\widehat{\text{LPML}} = \sum_{i=1}^{N} \text{log}(\widehat{\text{CPO}}_i).
\end{equation}
A larger LPML value indicates better model fit.

Similar to \cite{ma2019bayesianspatial}, we use the Rand index
\citep[RI;][]{rand1971objective} to evaluate the clustering performance,
i.e., whether the final inferred clusters align well with the truth.
Consider two partitions of $\{1,2,\ldots,n\}$, denoted as
$\mathcal{C}_1 = \{A_1,\ldots,A_r\}$ and~$\mathcal{C}_2 = \{
B_1,\ldots, B_s\}$. Out of all $n \choose 2$ pairs of observations, denote:
\begin{itemize}
    \item $a =$ the number of pairs that are in the same set in~$\mathcal{C}_1$
and in the same set in~$\mathcal{C}_2$
\item $b =$ the number of pairs that are in different sets in $\mathcal{C}_1$
and in different sets~in~$\mathcal{C}_2$
\item $c=$ the number of pairs that are in the same set in~$\mathcal{C}_1$ but
different sets~in~$\mathcal{C}_2$
\item $d=$ the number of pairs that are in different sets in~$\mathcal{C}_1$
but the same set set~in~$\mathcal{C}_2$.
\end{itemize}
With the above specifications, the RI is calculated as
\begin{equation}
    \mbox{RI} = (a+b) /(a+b+c+d) = (a+b) / \binom{n}{2}.
\end{equation}
It can be seen that the RI ranges from 0 to 1, with a larger value suggesting
better concordance between two clustering partitions.
Computation of the RI is done using the \textsf{R}
package \textbf{fossil} \citep{Rpkg:fossil}.

\section{Simulation}\label{sec:simu}
\subsection{Simulation Settings and Evaluation Metrics}

\begin{figure}[tbp]
	\includegraphics[width = \textwidth]{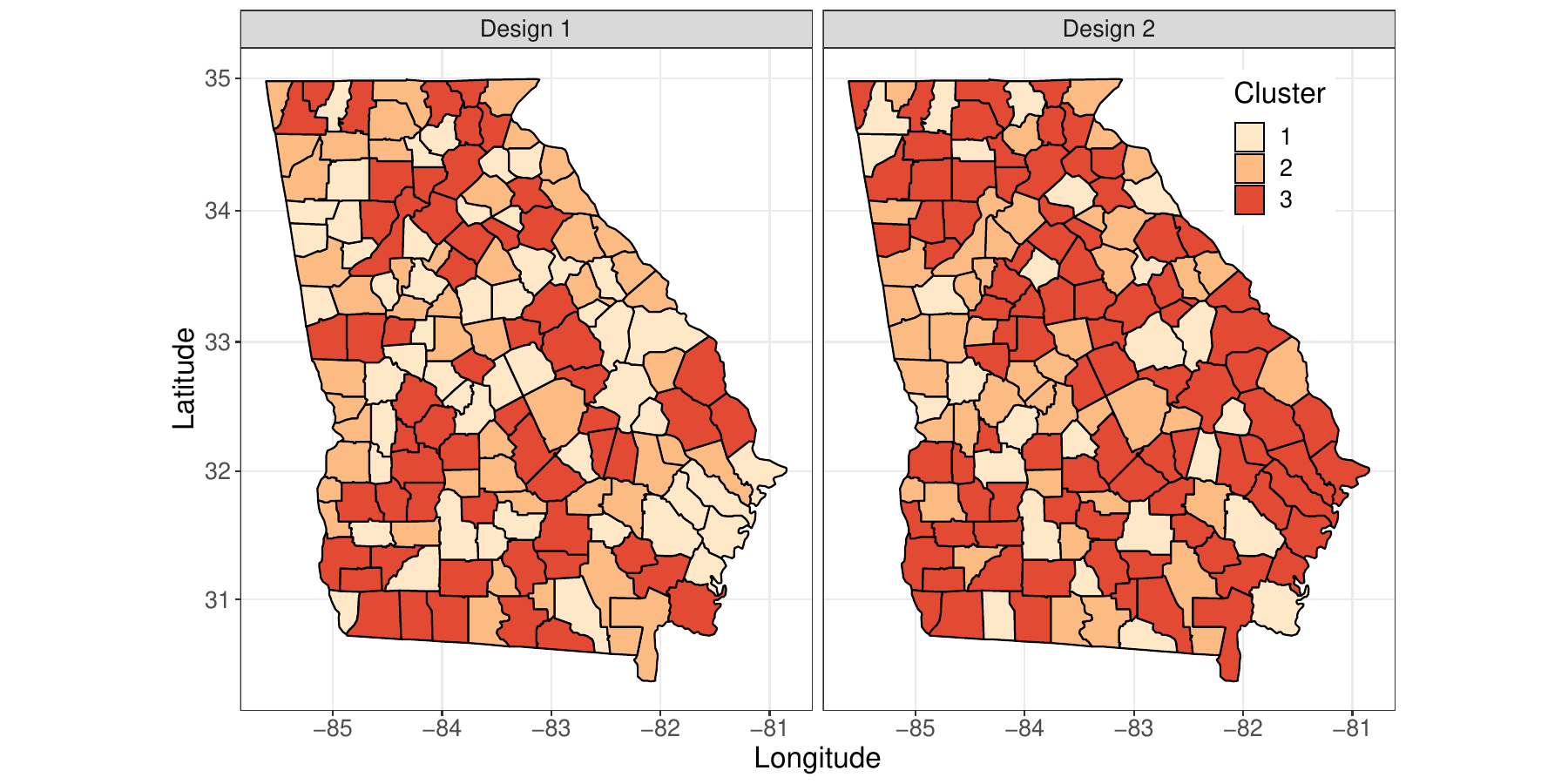}
	\caption{Visualization of simulation cluster designs. Design~1 has
balanced cluster sizes, while Design~2 has unbalanced cluster sizes.
	}\label{fig:simudesign}
\end{figure}

We study the estimation performance as well as the clustering performance in
this section. Two designs of true cluster configuration of Georgia counties are
considered. The first case is similar to in \cite{ma2019bayesianspatial}, where
there are, respectively, 51, 49, and 59 counties in each cluster. The second is
less balanced with 26, 44, and 89 counties in each cluster. The two partition
schemes used in designing the simulation study are visualized in
Figure~\ref{fig:simudesign}.

We consider the following data generation model:
\begin{equation}\label{eq:design3}
	\bm{Y} = \mbox{diag}(\bm{X} \bm{\beta}) + \bm{w} + \bm{\epsilon},
\end{equation}
where $\bm{w}$ is the distance- and auxiliary covariates-dependent vector of
spatial random effects such that
$$\bm{w}\sim \mbox{MVN}(\bm{0}, 0.25(0.81\bm{I} + 0.04\exp(-\mbox{GCD} / 4) +
0.05\bm{W}(\bm{Z}_1) + 0.1\bm{W}(\bm{Z}_2)),$$
with $\bm{Z}_1$ and $\bm{Z}_2$ being the two auxiliary covariates.

%
%
  The two similarity matrices $\bm{W}(\bm{Z}_1)$ and
  $\bm{W}(\bm{Z}_2)$ are constructed
using \eqref{eq:W} with $\kappa_1 =5$ and $\kappa_2 = 3$. The true parameters
for the similarity matrices are set to relatively small values compared to
$\alpha_0$ following \cite{zou2017covariance}. For both partition schemes shown
in Figure~\ref{fig:simudesign}, the true parameter vector for cluster~1 is set
to
$(4,1,-2)$, for cluster~2 $(1, 1, 0)$, and for cluster~3 $(1, -2, -1)$.
For each partition shown in Figure~\ref{fig:simudesign}, a total of~100
datasets are generated.

In addition to the proposed model, to verify that identifying clusters
do help with better estimation of the underlying coefficients, two additional
models are fitted.
The first alternative model is a Bayesian regression model either without
clusters or
spatial random effects, but includes the auxiliary covariates as 
main effects. It 
can be written hierarchically as
\begin{equation}
\begin{split}
	Y(s_i)\mid \bm{\beta} &\overset{\mbox{ind}}{\sim}
	\mbox{N}(\tilde{\bm{X}}(s_i)\bm{\beta},~ \tau_y^{-1}),\quad i=1,\ldots, 159,
\\
	\bm{\beta} & \sim \mbox{MVN}(\bm{0}, \sigma_\beta^2 \bm{I}_5),\\
	\tau_y &\sim \mbox{Gamma}(a, b),
\end{split}
\end{equation}
where $\tilde{\bm{X}}(s_i)$ in this case becomes
$\tilde{\bm{X}}(s_i) = 
\left(X_1(s_i), X_2(s_i), X_3(s_i), Z_1(s_i), Z_2(s_i)\right)^\top$,
and~$\bm{\beta}\in\mathbb{R}^5$.
We set~$\sigma_\beta^2=100$ to induce a non-informative prior for~$\bm{\beta}$.

The second alternative model is the Bayesian mixed model with spatial
random effects but without clustering, which can be written as
\begin{equation}
	\begin{split}
		Y(s_i) \mid \bm{\beta}, w_i & \sim \mbox{N}(\bm{X}(s_i)\bm{\beta} +
w_i,\tau_y^{-1}),\quad i=1,\ldots,159,\\
\bm{w} \mid \sigma^2 & \sim \mbox{MVN}(\bm{0}, \sigma^2 \exp(-\kappa \cdot
\mbox{GCD})),
\\
\bm{\beta} & \sim \mbox{MVN}(\bm{0}, \sigma_\beta^2
\bm{I}_3),\\
\sigma^2 &\sim \mbox{InverseGamma}(a_1,b_1),\\
\tau_y & \sim \mbox{Gamma}(c,d).
	\end{split}
\end{equation}
Again, the parameter~$\sigma_\beta^2$ is set to~100 to make a noninformative
prior for~$\bm{\beta}$.

%
%

The proposed approach and the two alternative models
are evaluated in terms of parameter estimation.
For estimation of the vector of coefficients, $\bm{\beta}$, we employ the mean
absolute bias (MAB), mean standard deviation (MSD), mean of mean squared
error (MMSE),
and mean coverage rate (MCR) for assessment:
\begin{align}
	\mbox{MAB} &= \frac{1}{159}\sum_{\ell=1}^{159}
	\frac{1}{100} \sum_{r=1}^{100}
	\left| \hat{\beta}_{\ell m r} - \beta_{\ell m}\right|, \\
	\mbox{MSD} & = \frac{1}{159}\sum_{\ell=1}^{159} \sqrt{\frac{1}{99}
\sum_{r=1}^{100}\left(\hat{\beta}_{\ell m r} -
\bar{\hat{\beta}}_{\ell m}\right)^2},
\\
\mbox{MMSE} & = \frac{1}{159} \sum_{\ell=1}^{159} \frac{1}{100}
\sum_{r=1}^{100}
\left(\hat{\beta}_{\ell m r} - \beta_{\ell m}\right)^2,\\
\mbox{MCR} & = \frac{1}{159}
\sum_{\ell=1}^{159}\frac{1}{100} \sum_{r=1}^{100} 1\left(\beta_{\ell m}
\in \mbox{HPD}_{\hat{\beta}_{\ell m r }}\right),
\end{align}
where $\hat{\beta}_{\ell m r}$ denotes the posterior estimate for the~$m$th
coefficient of county $\ell$ in the~$r$th replicate, $\bar{\hat{\beta}}_{\ell
m}
= \frac{1}{100}\sum_{r=1}^{100} \hat{\beta}_{\ell m r}$
, $\beta_{\ell m}$ is the
true underlying parameter value, $\mbox{HPD}_{{\hat{\beta}_{\ell m r
}}}$ is the 95\% highest posterior density interval for $\beta_{\ell m}$ in
the $r$th replicate, and $1(\cdot)$ denotes the indicator function.
Also, note that for the first alternative model, as we are primarily
interested in estimation of the three true main effects, we omit the
performance measures for the coefficients for the two auxiliary variables.

For each replicate, we set the chain length to 25,000 with thinning interval 2.
The first~9,500 of retained samples are discarded as burn-in, and we use the
remaining~3,000 iterations for posterior inference. The final cluster belonging
inferred for each county is taken as the first mode of the posterior samples
for
$z_i$, $i=1,\ldots, 159.$

\subsection{Simulation Results}

\begin{figure}[tbp]
	\centering
	\includegraphics[width = \textwidth]{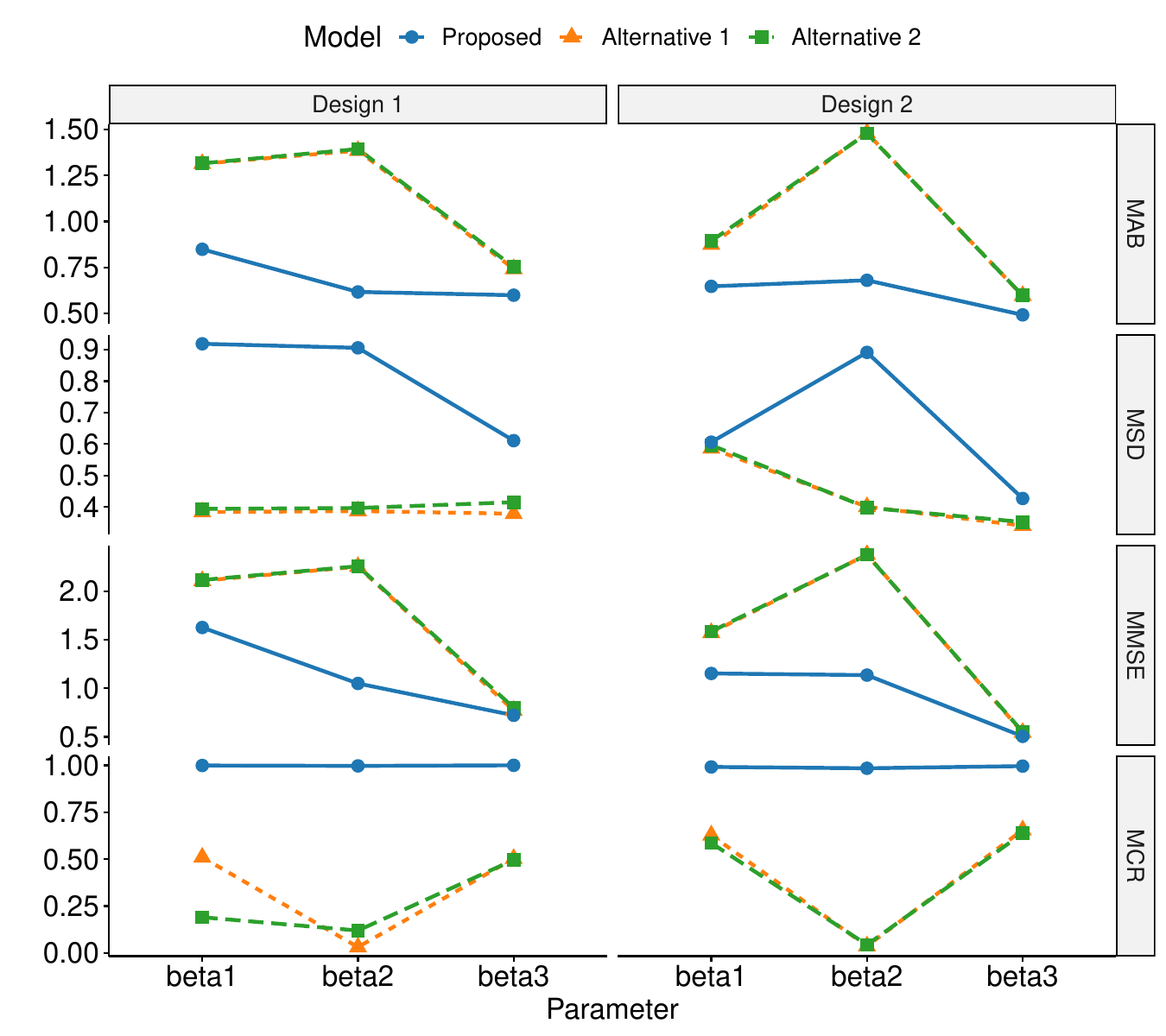}
	\caption{MAB, MSD, MMSE, and MCR of parameter estimates produced by the
produced model,
}
	\label{fig:performance_measure}
\end{figure}

First we check the estimation performance using the four performance
measures defined above.
For ease of reference, we name the three competitive models as
Alternative~1, Alternative~2, and proposed.
It can be seen from the first row that with the
incorporation of different clusters, each cluster of locations are allowed to 
have their own parameter vector. This additional flexibility
of the proposed model
enables less
biased parameter estimation. 
As the proposed model includes clustering
process, the chains for each parameter may jump between several underlying
clusters, which causes their MSD to be larger than those for Alternatives~1
and~2, which restrict that all locations have the same set of parameters.
However, with improved MAB, parameter
estimates produced by the proposed model
still have smaller MMSE than the
other two models. Finally, as Alternatives~1 and~2 do not allow for clusters
of coefficients, their parameter estimates are essentially close to the
average of parameters over the~159 locations, which leads to their very low
MCR.

\begin{figure}[tbp]
	\centering
	\includegraphics[width = \textwidth]{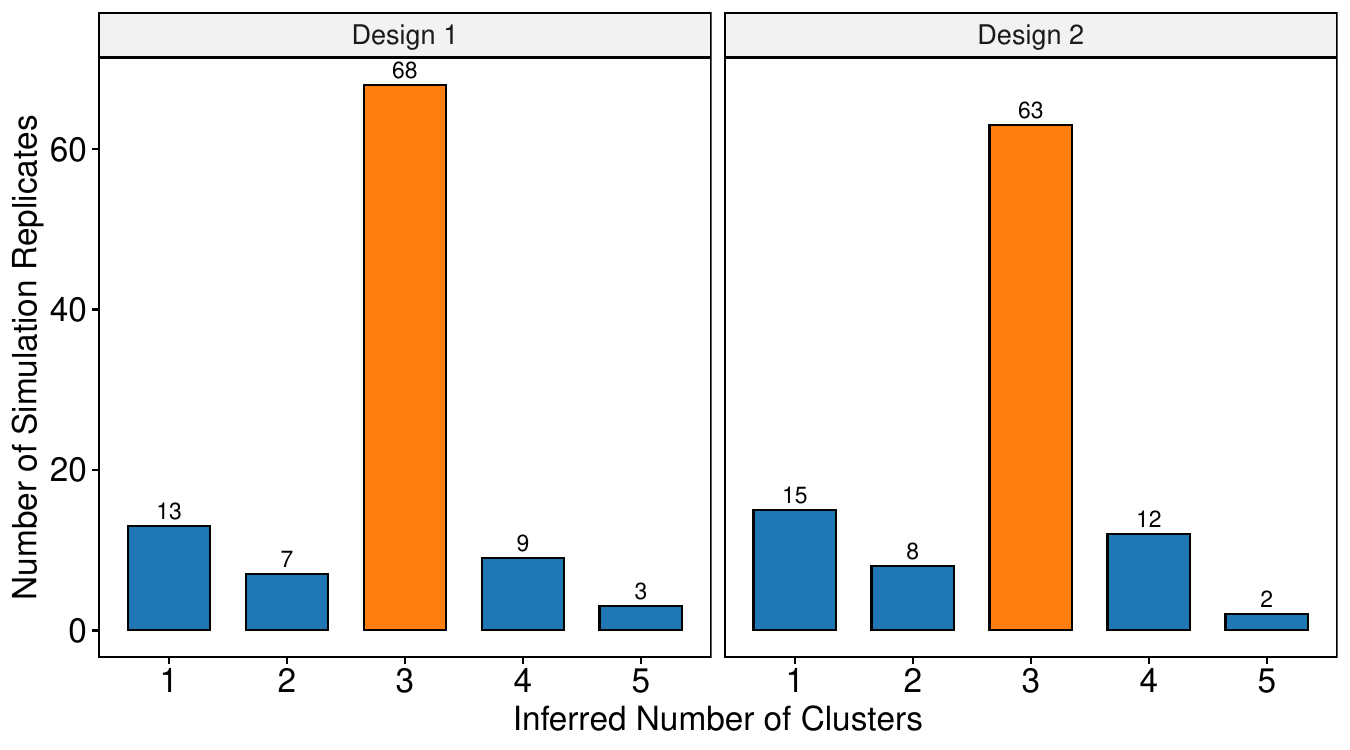}
	\caption{Histogram of clusters identified under the two true designs
	by the proposed model.
	The
orange bars correspond to the number of simulation replicates where the number
of clusters is correctly identified.}
	\label{fig:hist_K}
\end{figure}

The clustering performance of the proposed approach is presented in
Figure~\ref{fig:hist_K}. Comparing across the two panels, it can be seen that 
under Design~1 there are more replicates where~$K$ is correctly inferred,
while under Design~2 there are more under-clustering replicates, which is due
to its class imbalance. The average Rand index (ARI)'s turned out to be 0.703
and 0.752 for the two cases, respectively.

\begin{figure}[tbp]
    \centering
    \includegraphics[width=\textwidth]{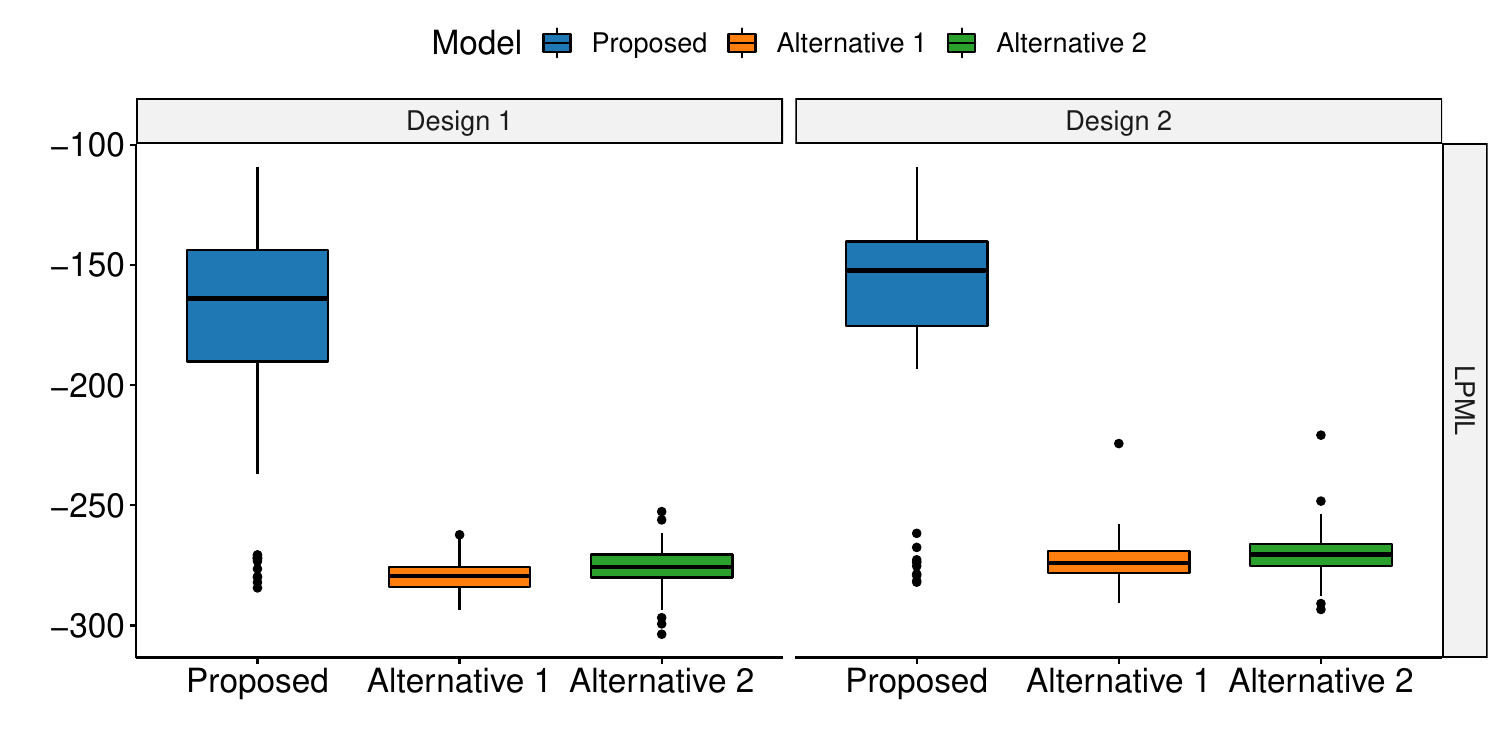}
    \caption{Boxplot of LPML values for each model.}
    \label{fig:LPML}
\end{figure}

Finally, to verify that LPML is capable of reflecting the degree of
fitness of the model to the data, for each simulation replicate, the LPML
values of the three models are calculated. A boxplot of the~100 LPML values
for each model under Designs~1 and~2 is given in Figure~\ref{fig:LPML}. As
discussed before, larger LPML values indicate better model fit. As clearly
seen in the figure, the proposed model has overall much larger LPML values
than the two alternatives, indicating that LPML is indeed capable of
identifying a more suitable model in the scope of the research problem
considered here.

\section{Real Data Analysis}\label{sec:real_data}
\subsection{Georgia Housing Cost Data}

The Georgia monthly housing dataset can be accessed at
\url{https://github.com/ys-xue/Bayesian-clustered-coefficients-regression
-ACAC} in \texttt{.csv} format. The original data source is
\url{www.healthanalytics.gatech.edu}, which contains visualizations of data
concerning multiple dimensions of Georgia.
For each of the 159 counties, the median
monthly housing cost for occupied housing units is observed. In addition,
several independent variables are available: the unemployment percentage for
adults between 18 and 64 years of age ($X_1$), the average per individual real
and personal property taxes ($X_2$), the median home market value in thousand
dollars ($X_3$), the White race population percentage ($Z_1$), the
median age ($Z_2$), and population size in thousands
($Z_3$).

In our analysis, the first three economy-related covariates, $X_1$, $X_2$ and
$X_3$, are used in the spatial regression part, while the remaining three
demographic covariates are used in constructing the covariance matrix of
spatial
random effects. The final model is written as, for $i=1,\cdots, 159$, 
\begin{equation*}
\begin{split}
        y(s_i) &\sim \mbox{N}(\mu_y(s_i), \tau_y^{-1}),\\
	\mu_y(s_i) &= \beta_0(s_i) +
	\beta_1(s_i) X_1(s_i) + \beta_2(s_i) X_2(s_i) + \beta_3(s_i)
X_3(s_i) + w(s_i),\\
	\beta_{z_i \ell} & \stackrel{\text{ind}} \sim \mbox{N}(\mu_{z_i \ell},
	\tau_{z_i \ell}^{-1}),
    ~~\ell = 1,\cdots, 3,\\
    z_i \mid k, \pi, \lambda & \sim \text{MFM}(\gamma, \lambda),\\
    \bm{w} &\sim \text{MVN}(\bm{0}, \bm{\Sigma}_{\bm{w}}),\\
	\bm{\Sigma}_{\bm{w}} & = \sigma^2\{ \alpha_0 \bm{I}_n + \alpha_1 W(\bm{Z}_1)
	+  \alpha_2
W(\bm{Z}_2) +  \alpha_3 W(\bm{Z}_3) +  \alpha_4 W(\bm{Z}_4)\},
\end{split}    
\end{equation*}
where the $(k,k')$-th element of $W(\bm{Z}_j)$ for $(j = 1, 2, 3)$ is
$\exp(- \kappa_j |Z_j(s_k) - Z_j(s_k')|)$,
respectively, while for $W(\bm{Z}_4)$, the entry is $\text{exp}(-\kappa_4 \cdot
\mbox{GCD})$. The priors of the unknown parameters are assigned as mentioned in
Section~\ref{sec:Infer}. Similar to in the simulation study, after burning in
the first 9,500 of 12,500 iterations, 3,000 MCMC samples are collected
the parameters.
Similar to in the simulation studies,
the final cluster configuration is obtained as the first
mode from the posterior samples in the chains corresponding
to~$z_1,\ldots,z_{159}$.

\subsection{Analysis Results}

We firstly apply the LPML to select the most suitable covariance structure of
spatial effects~$\bm{\Sigma}_{\bm{w}}$ for the model. The LPML values of the
proposed auxiliary covariates assistant covariance matrix, the unity scheme,
the
exponential scheme and the Gaussian scheme are shown in Table~\ref{tab:LPML}.
Comparison of the LPML values leads to the conclusion that the proposed
auxiliary covariates assisted covariance matrix provides the most suitable
approximation for the covariance structure of the spatial random effects for
this dataset, as it has the largest LPML value among the candidate covariance
structures. Therefore the auxiliary covariates assisted covariance matrix is
used in all subsequent analyses. The two alternative models we considered
in the simulation studies are also examined, and their LPML values are also
included in
Table~\ref{tab:LPML}.
Among the candidate models considered, the proposed model that employs the ACAC
has the largest LPML model, indicating that it is the most suitable choice to
capture the heterogeneity in the Georgia housing cost data.

\begin{table}
	\caption{\label{tab:LPML}LPML values for different models.}
	\centering
	\begin{tabular}{lcccccc}
		\toprule
& ACAC & Unity & Exponential & Gaussian & Alternative~1 &
Alternative~2 \\
		\midrule 
		LPML & -189.78 & -206.37 & -194.03 & -241.74 & -218.09  &
		-202.52\\
		\bottomrule
	\end{tabular}
\end{table}

Three clusters of the coefficients in the spatial regression
part~$\left\{\beta_\ell(\bm{s})\right\}_{\ell=1}^p$ are identified through the
MFM
approach, whose posterior estimates are shown in Table~\ref{tab:betaest} and
the
cluster belongings of the 159 counties are visualized in
Figure~\ref{fig:clusters}.
In addition, the traceplot for the number of clusters,~$k$, is included in the
supplemental material to verify convergence of the results. Convergence is
further verified with Dahl's method \citep{dahl2006model} in Section~S2 of the
supplemental material.
Cluster~1 includes~10 counties and cluster~3
consists
of 5 counties, while the rest~144 counties all fall within cluster~2. Taking a
closer look, cluster~1 consists of Fulton,
Douglas, 
Paulding, 
Henry, 
Newton, 
Barrow, 
Chattahoochee 
Lee,  
Effingham 
and Liberty, 
which are
all relatively economically developed counties in terms of per capita income
(among the top 50 according to 2015 United States Census Data and the 2006-2010
American Community Survey 5-Year Estimates) except Liberty. Cluster~3 consists
of Fannin,  
Union, 
Towns, 
Rabun, 
and Clay. 
Both clusters include neighboring counties and non-adjacent counties, which
again echos the finding in our simulation study that the proposed method takes
into consideration both spatial adjacency and the inherent similarity between
covariates that influence the spatial random effects.

\begin{table}
	\caption{\label{tab:betaest}Parameter estimates and their 95\% HPD intervals
for the three clusters identified.}
\centering
\begin{tabular}{lcccc}
	\toprule 
	Coefficient  & Cluster 1 &  Cluster 2 & Cluster 3  \\ 
	\midrule 
	$\hat{\beta}_0$ (Intercept) &   2.492 & -\textbf{4.943} & -0.801 \\
	& (-0.859, 5.288)&  (-\textbf{5.532}, -\textbf{4.252})   & (-3.521, 3.521)\\
	$\hat{\beta}_1$ (Unemployment Rate) & -0.280 & \textbf{0.578} & -1.011  \\
	& (-2.697, 2.996) & (\textbf{0.130}, \textbf{0.997}) & (-3.381, 1.197)   \\
	$\hat{\beta}_2$ (Tax) & -1.391 & -0.259 &  -0.441 \\
	& (-3.507, 0.583) & (-0.588, 0.047)  & (-1.633, 1.089)\\
	$\hat{\beta}_3$ (Home Market Value) &  1.526& \textbf{4.816} &  1.416 \\
	& (-0.417, 3.539)& (\textbf{4.444}, \textbf{5.177})  & (-0.821, 3.227) \\
	\bottomrule
\end{tabular}
\end{table}

\begin{figure}
	\centering
	\includegraphics[width = 0.8\textwidth]{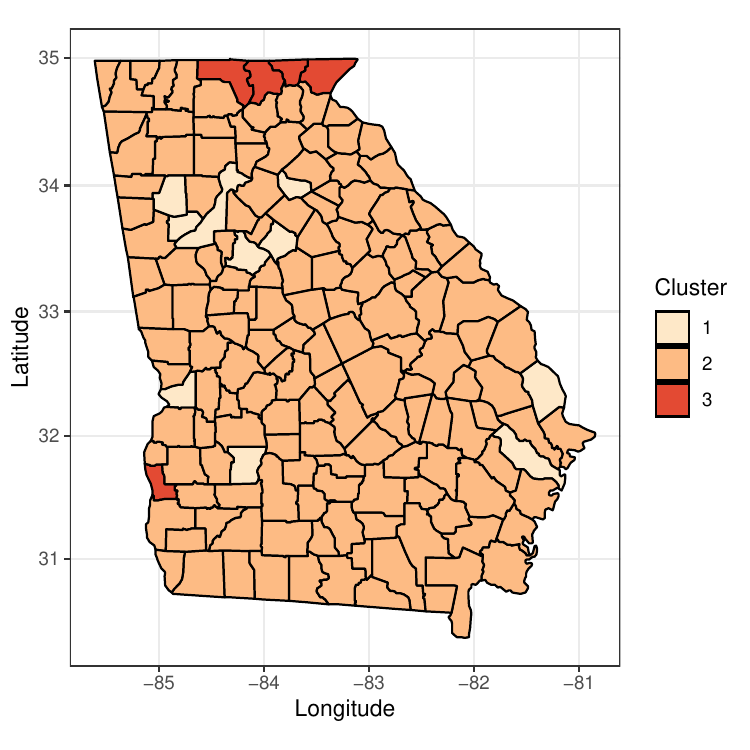}
	\caption{\label{fig:clusters}Clusters produced by the proposed approach.}
\end{figure}

From Table \ref{tab:betaest}, for counties belonging to cluster 2, the
percentage of unemployment and the median house market price can help explain
the change of median monthly housing cost.
However, for the other counties, neither the factors we selected has impact on
the dependent variable. Also, the intercept term for cluster~2 is noticeably
negative, indicating a difference in the overall level of housing cost between
counties in cluster~2 and those in the other two clusters.

Table~\ref{tab:otherest} shows the posterior estimates of the overall variance
term of spatial random effects, $\sigma^2$, and coefficients for the similarity
matrices. By comparing the posterior estimates of $\alpha_j (j = 0, \cdots, 4)$
in the auxiliary covariates assisted covariance matrix of the spatial effects,
we can see that the similarity matrices defined by the size of population and
the percentage of White race population have greater impact on the covariance
matrix of spatial effects. 

\begin{table}
\caption{\label{tab:otherest}Posterior estimates and their 95\% HPD
intervals for other parameters.}
\centering
	\begin{tabular}{lccc}
		\toprule 
		Parameters & Posterior estimate & SD & 95\% HPD interval  \\ \midrule 
		$\sigma^2$ &0.430 & 0.141 & (0.230, 0.787)\\
		$\alpha_0$ & 0.011 & 0.001 & (0.001, 0.030)\\
		$\alpha_1$ (White percentage) & 0.286 & 0.151 & (0.062, 0.624)\\
		$\alpha_2$ (median age) & 0.173 & 0.133 & (0.014, 0.515)\\
		$\alpha_3$ (population size) & 0.516 & 0.165 & (0.208, 0.819)\\
		$\alpha_4$ (GCD) & 0.014 & 0.012 & (0.000, 0.045)\\
		$\tau_y$ &  2.750 & 0.499 & (1.080, 3.818) \\
		\bottomrule
	\end{tabular}
\end{table}

\section{Discussion}\label{sec:discussion}

In this paper, we propose a Bayesian clustered coefficients regression model
with auxiliary covariates assistant random effects. Our proposed model has two
practical merits. First, our model simultaneously estimates the number of
clusters and clustering configurations of regression coefficients. Second,
auxiliary covariates information are included in our random effects model. The
usage of proposed method is illustrated in simulation studies, where it shows
accurate estimation and clustering performance. For Georgia housing cost data,
our method dominates the other benchmark methods in terms of LPML.

In addition, three topics beyond the scope of this paper are worth further
investigation. First, in our real data application, auxiliary covariates are
selected based on their natures, which is not always available or clearly
categorized in all possible applications. Proposing a quantitative criterion
for
auxiliary covariates determination is an interesting future work. Furthermore,
different clusters may have different sparsity patterns of the covariates.
Incorporating different sparsity structure of regression coefficients into the
model will enable selection and identification of most important covariates.
Finally, considering geographical information for clustering detection
\citep{hu2020bayesian,zhao2020bayesian,geng2020bayesian} is also devoted to
future research.

\end{document}